\documentclass[11pt]{article}
\usepackage{amssymb, amsmath, amsthm, amsfonts, amscd}
\usepackage[english]{babel}
\usepackage{graphicx}
\usepackage{caption}
\usepackage{subcaption}
\usepackage{color,soul}

\numberwithin{equation}{section}

\def\R{\mathbb{R}}

\newcommand{\set}[1]{\{ #1 \}}

\title{A penalization method for calculating the flow beneath travelling water waves of large amplitude}
\author{A.~Constantin\thanks{Faculty of Mathematics, University of Vienna, Oskar-Morgenstern-Platz 1, 1090 Wien, Austria} \and
K.~Kalimeris\thanks{Radon Institute of Computational and Applied Mathematics, Altenberger Str.~69, 4040 Linz, Austria} \and 
O.~Scherzer \thanks{Computational Science Center, University of Vienna, Oskar-Morgenstern-Platz~1, 1090 Wien, Austria and 
Radon Institute of Computational and Applied Mathematics, Altenberger Str.~69, 4040 Linz, Austria}
}

\begin{document}

\maketitle

\begin{abstract}
A penalization method for a suitable reformulation of the governing equations as a 
constrained optimization problem provides accurate numerical simulations 
for large-amplitude travelling water waves in irrotational flows and in flows with constant vorticity.
\end{abstract}

\section{Introduction}

Water flows with a uniform underlying current (possibly absent) are termed irrotational flows, while 
rotational waves describe the interaction of surface water waves with non-uniform currents. The study 
of the flow beneath an irrotational two-dimensional surface wave in water with a flat bed is quite 
well-understood: see \cite{C, CSp} for theoretical studies, \cite{Cl, Nach} for numerical simulations and 
\cite{Chen, U} for experimental data. For rotational two-dimensional travelling water waves an existence 
theory for waves of large amplitude is available \cite{CS} and some numerical simulations were performed 
in the case of constant vorticity flows without stagnation points \cite{KS1, KS2} and in the presence of stagnation points \cite{VO}. Constant non-zero vorticity is the 
hallmark of tidal currents, cf. the discussion in \cite{Cb}, and the absence of stagnation points excludes 
the possibility of a flow-reversal. These flows represent significant examples of rotational waves and 
our purpose is to pursue their in-depth study. We present a penalization method that selects from 
the family of solutions to a reformulation of the governing equations genuine
waves. This permits us to provide accurate simulations of the surface water wave 
but also of the main flow characteristics (fluid velocity components, pressure) beneath it.

\section{Preliminaries}

In this section we present the governing equations for periodic travelling water waves in a flow of 
constant vorticity over a flat bed. We briefly discuss the reformulation from \cite{CS} that leads, by 
means of bifurcation theory, to the existence of waves of small and large amplitude.

\subsection{Steady two-dimensional water waves}

Let us first discuss the governing equations for two-dimensional waves travelling at constant speed and 
without change of shape at the surface of a layer of water above a flat bed, in a flow of constant vorticity. 
Two-dimensionality means that the waves propagate in a fixed horizontal direction, say $X$, and the flow presents 
no variation in the horizontal direction orthogonal to the direction of wave propagation. For this reason, it 
suffices to analyse a vertical cross-section of the flow, parallel to the direction of wave propagation. To model 
sea waves of large amplitude the assumptions of inviscid flow in a fluid of constant density 
are appropriate and the effects of surface tension are negligible -- see the discussion in \cite{Cb}. The 
assumption of a flat bed $Y=-d$ is also reasonable for a considerable proportion of the Earth's sea floor. Consequently, 
the cross-section of the fluid domain is of the form
\begin{equation*}
\mathcal{D}(t) = \set{(X,Y) : X \in \R \text{ and } -d < y < \xi(X-ct)}\;,
\end{equation*} 
where $c>0$ is the wave speed, $d>0$ is the average depth and $\xi$ is the free surface. Setting the density of the water $\rho \equiv 1$, 
the incompressible Euler equations for the velocity field $(U(X-ct,Y),\,V(X-ct,Y))$ and the pressure $P(X-ct,Y)$ are
\begin{equation}
\left\{\begin{array}{l}
\label{Euler-eqs-old}
U_X+V_Y = 0\,, \\
(U-c)U_X + VU_Y = - P_X\,, \\
(U-c)V_X + VV_Y=-P_Y -g\,,
\end{array}\right. \quad \text{ in } \mathcal{D}(t)\,,
\end{equation}
where $g$ is the gravitational constant of acceleration. Since the flow is periodic in the $X$-variable, we may assume 
that the period is $2\pi$, after performing the rescaling $X \mapsto \frac{L}{2\pi} \,X$ in terms of the actual wavelength $L$. 
In a frame moving at the (constant) wave speed, obtained by means of the change of variables 
$$x=X-ct,\quad Y=y\,,$$ 
we can restrict our attention to the two-dimensional bounded domain
\begin{equation*}
\mathcal{D} = \set{(x,y): -\pi < x < \pi \text{ and } -d < y < \eta(x)}\,,\end{equation*} 
bounded above by the free surface profile
\begin{equation*}S = \set{(x,y): -\pi < x < \pi \text{ and }  y = \eta(x)}\,,\end{equation*} 
and below by the flat bed
\begin{equation*}B = \set{(x,y): -\pi < x < \pi \text{ and } y=-d}\,.\end{equation*}
Since $d$ represents the average depth, the waves oscillate around the flat free surface $y=0$, that is  
\begin{equation*} \int_{-\pi}^\pi\eta(x)\,dx=0,\end{equation*} 
where $\eta(x)=\xi(X-ct)$. Setting
$$u(x,y)=U(X-ct,y),\quad v(x,y)=V(X-ct,Y),\quad {\frak p}(x,y)=P(X-ct,Y),$$ 
(\ref{Euler-eqs-old}) can be written as
\begin{equation}
\left\{\begin{array}{l}
\label{ee}
u_x+v_y = 0, \\
(u-c)u_x+vu_y = - {\frak p}_x, \\
(u-c)v_x+vv_y=-{\frak p}_y -g\,,
\end{array}\right. \quad \text{ in } \mathcal{D}\,.
\end{equation}
The boundary conditions that select from the solutions to (\ref{ee}) those that represent water waves read as follows
\begin{equation}
\label{bc}
\left\{\begin{array}{l}
{\frak p} = P_{atm} \text{ on } S, \\
v = (u-c) \eta_x \text{ on } S,\\
v = 0 \text{ on } B\,,
\end{array}\right.\end{equation}
where $P_{atm}$ is the constant atmospheric pressure. The first condition 
reflects the fact that surface tension effects are negligible and permits the decoupling of the 
water motion from the air flow above it, while the second and third condition express the fact that the 
free surface and the flat bed are interfaces, with no flow possible across them -- see the discussion in \cite{Cb}. 

An essential flow characteristic is the vorticity $\gamma = v_x -u_y$, which is indicative of underlying currents. 
Vanishing vorticity is the hallmark of uniform currents and a constant vorticity characterizes the linearly sheared 
tidal currents. With respect to the flow beneath the waves, we restrict our attention to flows for which
\begin{equation}\label{ns}
u<c \quad\hbox{throughout the fluid}.
\end{equation}
This condition prevents the appearance of stagnation points in the flow and the occurrence of flow-reversals.

\subsection{Stream function formulation}

Structural properties of the governing equations (\ref{ee})-(\ref{bc}) enable us to reduce the number of 
unknowns. We first introduce the \emph{relative mass flux}\footnote{Relative to the uniform at speed $c$.}
\begin{equation}\label{mf}
p_0=\int_{-d}^{\eta(x)}\big(u(x,y)-c\big)\,dy<0,
\end{equation} 
since the first equation in (\ref{ee}) and the last two equations in (\ref{bc}) show that $\int\limits_{-d}^{\eta(x)}\big(u(x,y)-c\big)\,dy$ 
is independent of $x$, while (\ref{ns}) determines the sign. The first equation in (\ref{ee}) permits us to introduce the 
stream function $\psi(x,y)$ as the unique solution of the differential equations  
\begin{equation}
 \label{ds}
 \psi_x = -v, \qquad \psi_y = u-c \text{ in }\  \overline{\mathcal{D}}\,,
\end{equation}
subject to 
\begin{equation}
 \label{bs}
 \psi(x,-d) = -p_0\;.
\end{equation}
Note that $\psi(x,y)$ is periodic in the $x$-variable, and that the third equation in (\ref{bc}) is consistent with the 
constraint (\ref{bs}). Moreover, (\ref{ds}) and the definition of vorticity yield
\begin{equation}
\label{es}
\Delta \psi =-\omega\  \text{in }\  \mathcal{D}\;.
\end{equation}
The first equation in (\ref{bc}) is equivalent to $\psi$ being constant on $S$, while (\ref{mf}) together with (\ref{bs}) 
ensure that this constant must vanish, that is,
\begin{equation}
\label{ss}
\psi =0\ \text{ on } \ \mathcal{S}\;.
\end{equation}
On the other hand, due to (\ref{ds}), we see that we can re-express the Euler equation in (\ref{ee}) by the fact that 
the expression $\frac{(u-c)^2+v^2}{2}+gy+{\frak p}+\gamma\psi$ equals a constant $E$ throughout $\mathcal{D}$. 
The constant $Q=E-P_{atm}$ is called the hydraulic head. 

The previous considerations show that the governing equations (\ref{ee})-(\ref{bc}) can be reformulated in terms of the 
stream function as the free-boundary problem
\begin{equation}
\label{str}
\left\{\begin{array}{l}
\Delta\psi=-\omega\quad\text{in}\quad {\mathcal D}\,,\\
\psi=0 \quad\hbox{on}\quad S\,,\\
\psi=p_0 \quad\hbox{on}\quad B\,,\\
\frac{|\nabla\psi|^2}{2} + gy =Q\quad\hbox{on}\quad S\,.
\end{array}\right.
\end{equation}
Given $p_0$, we seek values of $d$ and $Q$ for which (\ref{str}) admits a smooth solution $\psi(x,y)$, even and of period $2\pi$ 
in the $x$-variable. Evenness reflects the requirement that $u$ and $\eta$ are symmetric while $v$ is antisymmetric about 
the crest line $x=0$; here, we shift the moving frame to ensure that the wave crest is located at $x=0$. Symmetric waves present 
these features and it is known that a solution with a free surface $S$ that is monotone between crest and trough has to be symmetric, 
cf. \cite{CEW}.

\subsection{Hodograph transform}

Under the assumption (\ref{ns}), a partial hodograph transform leads to a reformulation of the free-boundary 
problem \eqref{str} as a quasilinear elliptic system in a known strip. In this process, the wavelength is normalized to 
$2\pi$ and the gravitational constant $g$, 
the relative mass flux $p_0$ and the constant vorticity $\gamma$ are considered to be known, while the average depth $d$ 
and the hydraulic head $Q$ are allowed to vary to accommodate the existence of a flow.

The assumption (\ref{ns}) and the definition of the stream function \eqref{ds} yield that $\psi(x,y)$ is a strictly decreasing 
function of $y$ throughout the fluid domain ${\mathcal D}$, being periodic in the $x$-variable. Moreover, 
due to (\ref{bs}) and (\ref{ss}), $\psi$ is constant both 
on the bottom $B$ and on the free surface $S$. The Dubreil-Jacotin transformation \cite{DJ} 
\begin{equation*}
q=x, \qquad p=-\psi,
\end{equation*}
transforms the unknown domain $\mathcal{D}$ to the rectangle 
\begin{equation}
 R=\set{(q,p):\ -\pi<q<\pi\,,\  p_0<p<0}\;,
\end{equation}
\begin{figure}
\centering
\includegraphics[height = 45mm, width =95mm]{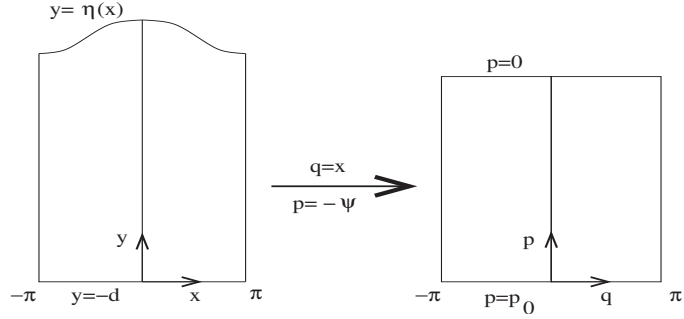}
\caption{Dubreil-Jacotin transformation}
\label{D-J}
\end{figure}
(see Figure 1). Let 
\begin{equation} \label{def:h} h(q,p) = y + d \end{equation} 
define the height above the flat bottom $B$. Since $\psi$ is a strictly decreasing function of $y$, for every fixed $x$ the height $h$ above the flat bottom is a single valued function of $\psi$ (or, equivalently, $p$), with
\begin{equation*}
\left\{\begin{array}{ccccc}
h_q &=& \dfrac{v}{u-c}, \quad h_p &=& \dfrac{1}{c-u}\,,\\[0.33cm]
v &=& -\dfrac{h_q}{h_p}, \quad u &=& c-\dfrac{1}{h_p}\,,
\end{array}\right.
\end{equation*}
and, more generally,
\begin{equation}\label{partial-rel}
\left\{\begin{array}{ccccc}
\partial_x &=& \partial_q -\dfrac{h_q}{h_p}\,\partial_p\,,\quad \partial_y &=& \dfrac{1}{h_p}\,\partial_p\,,\\[0.2cm]
\partial_p &=& \dfrac{1}{c-u}\partial_y \,,\quad \partial_q &=& \partial_x - \dfrac{v}{c-u} \partial_y\;.
\end{array}\right.
\end{equation}
Using the change of variables relations \eqref{partial-rel}, we get that
\begin{align*}
\gamma &= \partial_x v -\partial_y u= \left(\partial_q -\frac{h_q}{h_p}\partial_p \right) \left( -\frac{h_q}{h_p}\right) -\frac{1}{h_p}\partial_p\left( c-\frac{1}{h_p} \right) \\
&=\dfrac{-h_p h_{qq}+h_qh_{pq}}{h_p^2}-\dfrac{-h_ph_qh_{pq}+h_q^2 h_{pp}}{h_p^3}-\dfrac{h_{pp}}{h_p^3},
\end{align*}
while
\begin{equation*}|\nabla\psi|^2=v^2+(u-c)^2=\dfrac{1+h_q^2}{h_p^2}\,.\end{equation*}
These considerations show that the constitutive equations for the height function $h(q,p)$, which is even and $2\pi$-periodic in $q$, are 
\begin{equation}\label{op_basic}
\left\{\begin{array}{l}
\mathcal{H}[h] := (1+h_q^2)h_{pp} - 2h_ph_qh_{pq} +h_p^2h_{qq} +\gamma h_p^3 =0\  \text{ on } \ R\,,\\
\mathcal{B}_0[h] := 1+h_q^2(q,0) + (2gh-Q)h_p^2(q,0) = 0 \,, \\
\mathcal{B}_1[h] := h(q,p_0) = 0\,.
\end{array}\right.
\end{equation}
In the new formulation \eqref{op_basic}, the wave profile $\eta(x)$ is given by $h(q,0)$, the wave height being the difference 
\begin{equation}\label{def:amplitude}
\max_{q\in[-\pi,\pi]} h(q,0)- \min_{q\in[-\pi,\pi]} h(q,0).
\end{equation}
while half of \eqref{def:amplitude} represents the wave amplitude.

\section{Laminar Flow and Linearised Equations}

The simplest solutions are the laminar flows with a flat free surface. Near such flows a linearization procedure permits us to 
obtain the first-order approximations of genuine water waves. These linear waves capture well the characteristics of waves 
of small amplitude.

\subsection{Laminar flows}

Let us discuss the solutions describing parallel shear flows, with $\eta \equiv 0$. 
In this case the solution $h$ of \eqref{op_basic} is independent of $q$: $h(q,p)=H(p)$ with
\begin{equation}\label{eH}
\left\{\begin{array}{l}
\mathcal{H}_L[H] := H_{pp} + \gamma H_p^3 =0 \text{ in } R\,,\\
\mathcal{B}_{L,0}[H] := 1 + (2gH(0)-Q)H_p^2(0) = 0\,, \\
\mathcal{B}_{L,1}[H] := H(p_0) = 0\;.
\end{array}\right.
\end{equation}
The explicit solution of \eqref{eH} is given by
\begin{equation} \label{H}
H(p;\lambda) = \frac{2(p-p_0)}{\sqrt{\lambda + 2 \gamma p}+\sqrt{\lambda+2\gamma p_0}}\,,\quad p_0 \le p \le 0\,,
\end{equation}
provided that the parameter $\lambda>0$ satisfies the equation
\begin{equation} \label{Q}
Q = \lambda + \frac{4 g |p_0|}{\sqrt{\lambda} + \sqrt{\lambda+ 2 \gamma p_0}}\,.
\end{equation} 

\subsection{Linearised Solutions}

We now present the outcome of the linearization of the system \eqref{op_basic} near the laminar flow $H$. 

We consider a parametrized family of functions of the form 
\begin{equation}
\label{eq:hat_h}
\hat{h}(q,p) = H(p) + b m(q,p) \,,
\end{equation}
where $b \in \R$ and the function $m$ is even and $2\pi$-periodic in $q$, such that 
\begin{equation}
\label{eq:ob2}
 \mathcal{H}[\hat{h}](p,q) = \mathcal{O}(b^2)\,,\; \mathcal{B}_0[\hat{h}](q) = \mathcal{O}(b^2) \text{ and } \mathcal{B}_1[\hat{h}](q) = 0\;.
\end{equation}
Taking the definition of $\mathcal{H}, \mathcal{B}_0, \mathcal{B}_1$ from \eqref{op_basic} into account, we find that 
$\mathcal{H}[\hat{h}](p,q)$ is given by
\begin{equation*}
 \begin{aligned}
 & (1 + H_q^2(p) + 2b H_q(p) m_q(q,p) + b^2 m_q(q,p)) (H_{pp}(p) + b m_{pp}(p,q))\\
   & - 2(H_p(p)+b m_p(q,p))(H_q(p)+bm_q(q,p))(H_{pq}(p)+b m_{pq}(q,p)) \\
   & + (H_p^2(p)+2b H_p(p) m_p(p,q) +b^2m_p^2(p,q)) (H_{qq}(p)+bm_{qq}(p,q)) \\
   & + \gamma (H_p^3(p) + 3 b m_p(p,q) H_p(p) + \mathcal{O}(b^2))\;.
 \end{aligned}
\end{equation*}
Using the fact that $H_q = 0$, the expression for $\mathcal{H}[\hat{h}](p,q)$ simplifies to 
\begin{equation*}
\begin{aligned}
 & (1 + b^2 m_q(q,p)) (H_{pp}(p) + b m_{pp}(p,q))\\
   & - 2 b^2 (H_p(p)+b m_p(q,p) m_q(q,p) m_{pq}(q,p) \\
   & + b (H_p^2(p)+2b H_p(p) m_p(p,q) +b^2m_p^2(p,q)) m_{qq}(p,q) \\
   & + \gamma (H_p^3(p) + 3 b m_p(p,q) H_p^2(p) + \mathcal{O}(b^2))\\
  &= H_{pp}(p) + \gamma H_p^3(p)\\
   & + b (m_{pp}(p,q)+H_p^2(p)m_{qq}(p,q) + 3 \gamma m_p(p,q) H_p^2(p))  + \mathcal{O}(b^2)\;.
 \end{aligned}
\end{equation*}
Similarly,
\begin{equation*}
 \begin{aligned}
 ~ & \mathcal{B}_0[\hat{h}](q) =1 + (2gH(0)-Q)H_p^2(0) \\
   & + b 2H_p(0) \left((2gH(0)-Q) m_{p}(q,0)+ g H_p(0) m(q,0) \right)  + \mathcal{O}(b^2)\;.
 \end{aligned}
\end{equation*}
Using the fact that \eqref{eH} and \eqref{H} yield $H_p(0)=\frac{1}{\sqrt{\lambda}}$ and $2gH(0)-Q=-\lambda$, we infer that
\begin{equation*}
 \begin{aligned}
 ~ & \mathcal{B}_0[\hat{h}](q) =1 + (2gH(0)-Q)H_p^2(0) \\
   & + b \frac{2}{\lambda} \left( -\lambda^{3/2} m_{p}(q,0)+ g m(q,0) \right)  + \mathcal{O}(b^2)\;.
 \end{aligned}
\end{equation*}
$H$ being a solution to \eqref{eH} shows that $\hat{h}$ solves \eqref{eq:ob2} if and only if $m$ satisfies the linearised system
\begin{equation}
\label{linear_eq_aux}
\left\{\begin{array}{l}
m_{pp} + H_p^2 m_{qq} = -3 \gamma H_p^2 m_p \text{ in } R\,, \\ 
g m(q,0) = \lambda^{3/2} m_p(q,0) \text{ for } -\pi < q < \pi\,,\\
m(q,p_0) = 0 \text{ for } -\pi < q < \pi\;.
\end{array}\right.
\end{equation}

For a general value of $\lambda>0$, the problem \eqref{linear_eq_aux} will admit only the trivial solution $m \equiv 0$. 
However, specific values of $\lambda$ produce non-trivial solutions:
\begin{itemize}
\item In the irrotational case $\gamma \equiv 0$ we have that 
the solution to \eqref{eH} is
\begin{equation}
H(p;\lambda)=\dfrac{p-p_0}{\sqrt{ \lambda }}\,,
\end{equation}
and the non-trivial solution of the linearized equations \eqref{linear_eq_aux} is given by $m(q,p)=M(p)\,\cos(q)$ with
\begin{equation}
M(p) = \sinh\left( \dfrac{p-p_0}{\sqrt{ \lambda^* }}\right),
\end{equation}
where $\lambda^*>0$ satisfies the dispersion relation
\begin{equation}
\lambda + g \tanh \left( \dfrac{p_0}{\sqrt{ \lambda }}\right) = 0\,,
\end{equation}
the corresponding value of $Q$ being 
\begin{equation*}Q^* = \lambda^* - \dfrac{2 g p_0}{\sqrt{ \lambda^*}}\;.\end{equation*}
We obtain the linear solution
\begin{equation}
\label{sol_perturb_ir}
h^*(q,p;b)= \dfrac{p-p_0}{\sqrt{ \lambda^* }} + b \cos q \ \sinh\left( \dfrac{p-p_0}{\sqrt{ \lambda^* }}\right), 
\end{equation}
with $b$ constant. This is the first-order approximation to the solution of the problem 
\eqref{op_basic}, up to order $\mathcal{O}(b^2)$, for small enough $b$.
\item Similarly, in the case of constant non-zero vorticity $\gamma$, for  
\begin{equation*}Q^*= \lambda^* -\dfrac{4 g p_0}{\sqrt{ \lambda^* } +\sqrt{ \lambda^* +2 p_0 \gamma }} \end{equation*} 
we get the linear solution 
\begin{equation}
\label{sol_perturb}
h^*(q,p;b)= H^*(p) + b \cos q \ M(p), 
\end{equation}
with
\begin{equation}
H^*(p) = \frac{2(p-p_0)}{\sqrt{\lambda^* + 2 \gamma p}+\sqrt{\lambda^*+2\gamma p_0}}
\end{equation}
and
\begin{equation}
M(p) = \dfrac{1}{\sqrt{ \lambda^* +2 p \gamma }}\sinh\left(  \frac{2(p-p_0)}{\sqrt{\lambda^* + 2 \gamma p}+\sqrt{\lambda^*+2\gamma p_0}}\right),
\end{equation}
where $ \lambda^*>0$ is the solution of the dispersion relation
\begin{equation}\label{disp}
\dfrac{ \lambda }{g+ \gamma \sqrt{ \lambda }} + \tanh \left( \dfrac{2p_0}{ \sqrt{ \lambda } +\sqrt{ \lambda +2 p_0 \gamma } }\right) =0 .
\end{equation}
\end{itemize}

\subsection{Bifurcation}

The interpretation of the previous results in the space of solutions is provided by means of bifurcation theory: 
near the laminar flows \eqref{H}, as the parameter $\lambda$ varies, there are generally no genuine waves, except at critical 
values $\lambda=\lambda^\ast$ determined by the dispersion relation \eqref{disp}. Note that by \eqref{H} and \eqref{partial-rel}, 
we have that $\sqrt{\lambda}=\frac{1}{H_p(0;\lambda)}=c-u(0,0)$, so that this result means that only critical values of 
the horizontal fluid velocity of the laminar flows at their flat free surface may trigger the appearance of waves. Near this 
bifurcating laminar flow $H^\ast$, we have two solution curves: one laminar solution curve $\lambda \mapsto H(p;\lambda)$, where $\lambda$ and $Q$ are related by \eqref{Q}, and one non-laminar solution curve $Q \mapsto h(q,p;Q)$ such that $h_q \not \equiv 0$ unless $h=H^\ast$. 
In \cite{CS} it was shown that non-laminar solution curve can be extended to a global continuum $\mathcal{C}$ that 
contains solutions of \eqref{op_basic} with $\frac{1}{h_p(q_0,p_0)}\to 0$ at some $(q_0,p_0)$. This condition is characteristic of flows 
such that their horizontal velocity $u$ is arbitrarily close to the speed $c$ of the reference frame, at some point in the fluid, 
the limiting configuration being a flow with stagnation points.

\section{Optimization}
In the following, we consider the numerical solution of the free boundary value problem for water waves with constant vorticity.

We propose a \emph{Penalization Method} (PM) for solving the constraint optimization problem, to minimize 
\begin{equation}
 \label{eq:energy}
 \mathcal{E}[h]:=-\int_{R} h_q^2\,,
\end{equation}
subject to the PDE constraint that $h$ satisfies \eqref{op_basic}.

The energy function $\mathcal{E}$ is chosen in such a way that it vanishes for laminar flows (in which $h_q\equiv 0$), thus selecting genuine waves.

We propose the following implementation the PM method: 
\begin{enumerate}
 \item Initialize $k=0$: Choose a constant $\nu_0 > 0$ (typically small). Find an initial guess $h^{(0)}$ of the solution of \eqref{op_basic}. 
       For initializing $h^{(0)}$ we select $h^{(0)}(q,p) = h^*(q,p;b)$  which gets the particular forms \eqref{sol_perturb_ir} and \eqref{sol_perturb}, for $\gamma=0$ and $\gamma\neq 0$, respectively.  These forms guarantee that \eqref{eq:ob2} holds.
       
       In order to calculate 
       waves of large amplitude (one branch of the bifurcation is the laminar flow, and the other branch is the one with high amplitude), 
the particular choice of $b$ is important for initialization: On the one hand the closer $b$ is to $0$, the smaller the residual is (cf. \eqref{eq:ob2} ). 
       On the other hand for $b=0$, $h^*(q,p;0)$ is a laminar flow, and the PM algorithm is attracted to the laminar flow solution.

 \item  $k \to k+1$: Given $h^{(k)}$ we solve the following linear equation for $h$, obtained by freezing the coefficients of lower order from the previous iteration step
       \begin{equation}
       \label{basic_eq_num}
       \begin{aligned}
       0 = & \mathcal{H}_l[h^{(k)}](h) \\
         = & (1+(h_q^{(k)})^2)h_{pp} - 2h_p^{(k)}h_q^{(k)}h_{pq} +(h_p^{(k)})^2h_{qq} + \gamma \ (h_p^{(k)})^3\\
           & \qquad \text{ in } R,\\
       0 = & \mathcal{B}_{l,1}[h^{(k)}](h) \\
         = & 1+(h_q^{(k)})^2 + (2gh-Q)(h_p^{(k)})^2 \text{ for } p=0,\\
       0 = & \mathcal{B}_{l,2}[h^{(k)}](h) = h \text{ for }p=p_0\;.
       \end{aligned}
       \end{equation}
       The solution is denoted by $h^{(k+1)}$.
\item
Compute $h_p^{(k+1)}$. Because we work with a semi implicit scheme we have to use a relative small step-size, which is determined here.
\begin{itemize} 
\item If $h_p^{(k+1)}>0$ then put $\nu_{k+1} = \nu_k$ and update $  h^{(k+2)}(q,p) = h^{(k+1)}(q,p) + \nu_{k+1} h^{(k+1)}_{qq}(q,p).$ We emphasize that $h^{(k+1)}_{qq}$ is the steepest descent energy
      of the quadratic functional $\mathcal{E}$. From this perspective we might call this algorithm a \emph{steepest descent} algorithm. 

\item else put $\nu_{k+1}=0$ and  update  $$ h^{(k+2)}(q,p)  = F(p) - h^{(k)}(q,p).$$ The function $F$ is given by  $F(p) \simeq 2 \frac{d}{d^*} H^*(p)$ 
with $d^*$ and $d$ being the depths of $H^*(p)$ and $h^{(k)}(q,p)$, respectively. Using \eqref{def:h} we see that the depth $d$ of a flow $h(q,p)$ can be computed by 
\begin{equation*}
d=\int_{-\pi}^\pi h(q,0)dq.
\end{equation*} 
\end{itemize} 
\item Stopping criteria: The algorithm is terminated when the system of equations \eqref{op_basic} is satisfied up to small error, i.e., 
\begin{itemize}
\item $||\mathcal{H}_l[h^{(k)}](h^{(k)})||<\epsilon_1$ and $||\mathcal{B}_{l,1}[h^{(k)}](h^{(k)})||<\epsilon_2$
\end{itemize}
and
    \begin{itemize}
\item $||h^{(k)} -h^{(k-1)}||<\epsilon_3$, or
\item $0<1/h^{(k)}_p<\epsilon_4$, which means that we are close to a stagnation point, i.e. $c-u = 1/h_p$ is positive and close to zero.
\end{itemize}
If the algorithm is not terminated then move to the second step.
\end{enumerate}

\begin{figure}
\centering
\begin{subfigure}{.5\textwidth}
  \centering
  \includegraphics[width=1\linewidth]{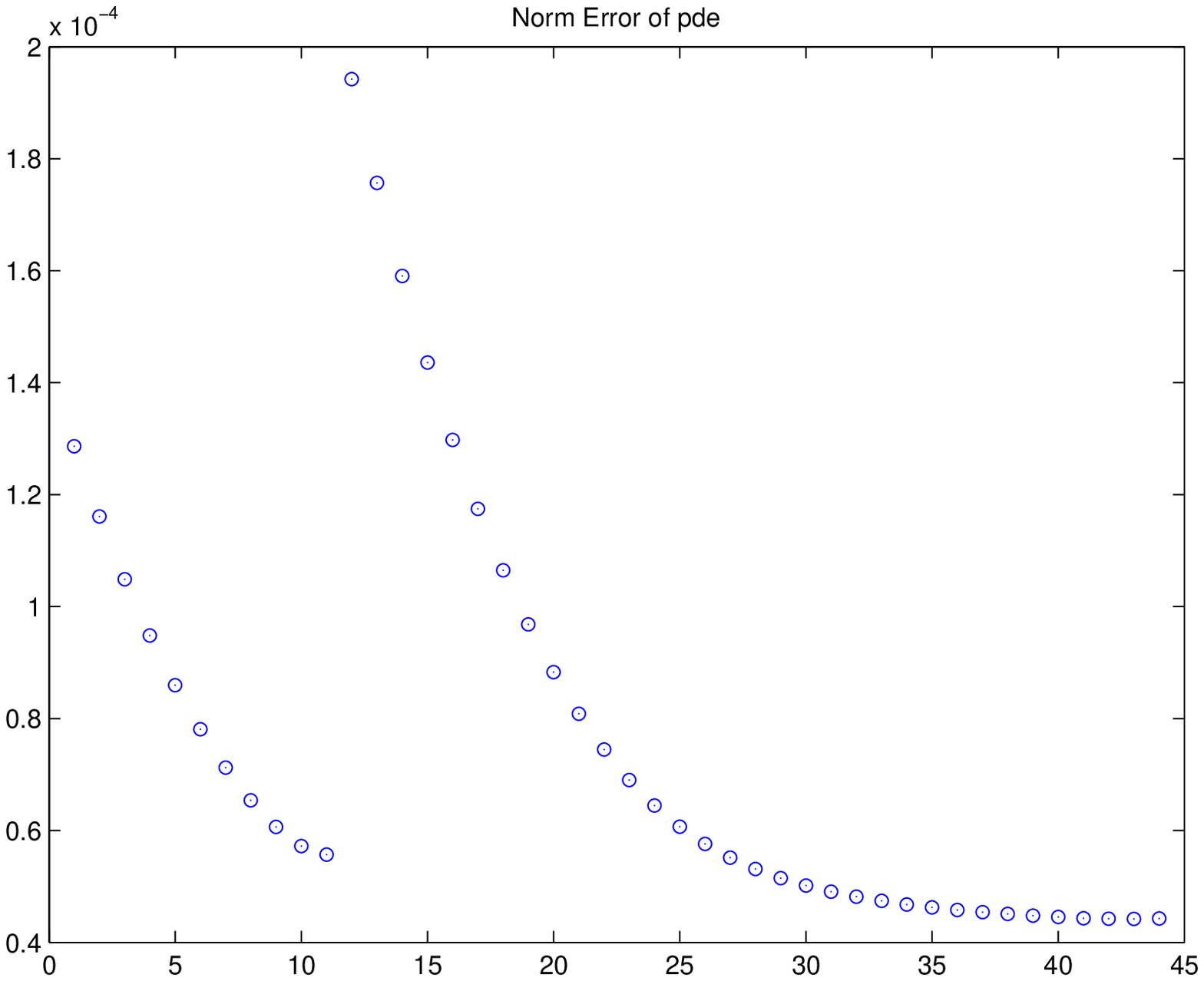}
  \caption{The error of the PDE.}
\end{subfigure}%
\begin{subfigure}{.5\textwidth}
  \centering
  \includegraphics[width=1\linewidth]{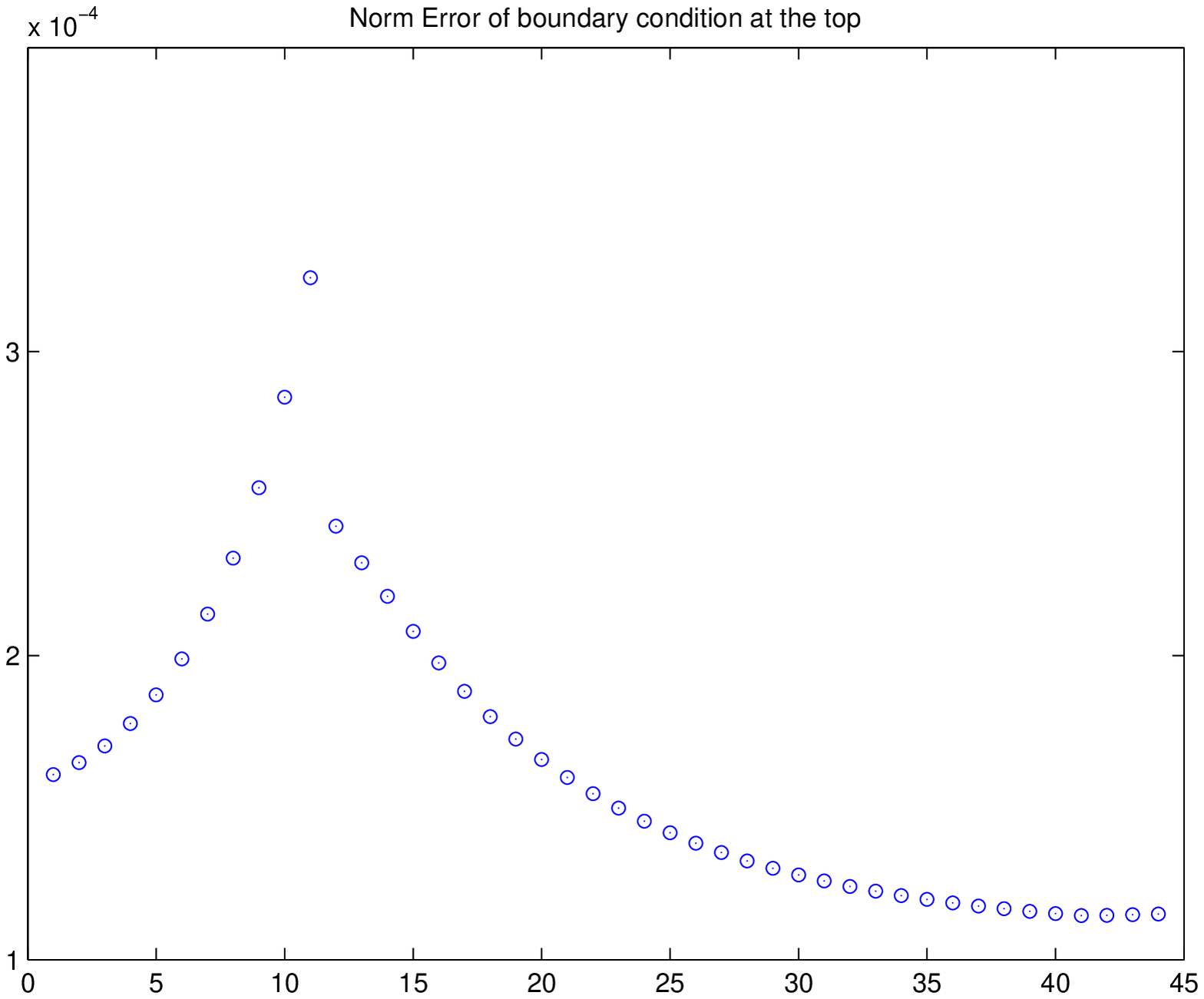}
  \caption{The error of the Boundary condition.}
\end{subfigure}
\caption{The $L_2$ norms of the errors for each iteration.}
\label{fig-er}
\end{figure}

The different branches of the third step guarantee that the residuals of the boundary conditions and the differential equations are decreasing. We have illustrated this for one example in Figure \ref{fig-er}, 
where the necessity of changing the iteration becomes evident. More precisely, for the above algorithm, there is an iteration $m$ that the condition $h_p^{(m+1)}>0$ fails and we make the update 
$  h^{(m+2)}(q,p) = F(p) - h^{(m)}(q,p)$.

The simulations show  that for all $k=1,\ldots,m$ \begin{equation*} ||\mathcal{H}_l[h^{(k)}](h^{(k)})||<||\mathcal{H}_l[h^{(k-1)}](h^{(k-1)})|| \end{equation*} and 
\begin{equation*} 
||\mathcal{B}_{l,1}[h^{(k)}](h^{(k)})||>||\mathcal{B}_{l,1}[h^{(k-1)}](h^{(k-1)})||,
\end{equation*} as this is also depicted in Figure \ref{fig-er}. For $k \geq m+2$ we find that \begin{equation*} ||\mathcal{H}_l[h^{(k+1)}](h^{(k+1)})||<||\mathcal{H}_l[h^{(k)}](h^{(k)})|| \end{equation*} and \begin{equation*} ||\mathcal{B}_{l,1}[h^{(k+1)}](h^{(k+1)})||<||\mathcal{B}_{l,1}[h^{(k)}](h^{(k)})||,\end{equation*} as depicted in Figure \ref{fig-er}.

\section{Results}

We use that the gravitational constant $g=9.8$ and we fix the relative mass flux $p_0=-2$. 
We present the results of the numerical simulations for three different cases:
\begin{itemize}
\item the irrotational case, $\gamma=0$,
\item the case of positive vorticity $\gamma=2.95$,
\item the case of negative vorticity $\gamma=-1$.
\end{itemize}
For all the cases we present the figures of 
\begin{itemize}
\item the profile of the water wave, that is  the free boundary $S$, 
\item the streamline pattern beneath the wave, that is  the height $h(q,p)$ along the streamlines $p=-\psi$.
\item the distribution of the vertical velocity $v$ in the whole fluid.
\item the horizontal velocity $c-u$ on the vertical line below the crest, that is  on the straight line segment $\{(q,p): \ q=0, \ p\in[p_0,0]\}$.
\item the distribution of the pressure beneath the wave.
\end{itemize}

\begin{figure}
\centering
\begin{subfigure}{.33\textwidth}
  \centering
  \includegraphics[width=1\linewidth]{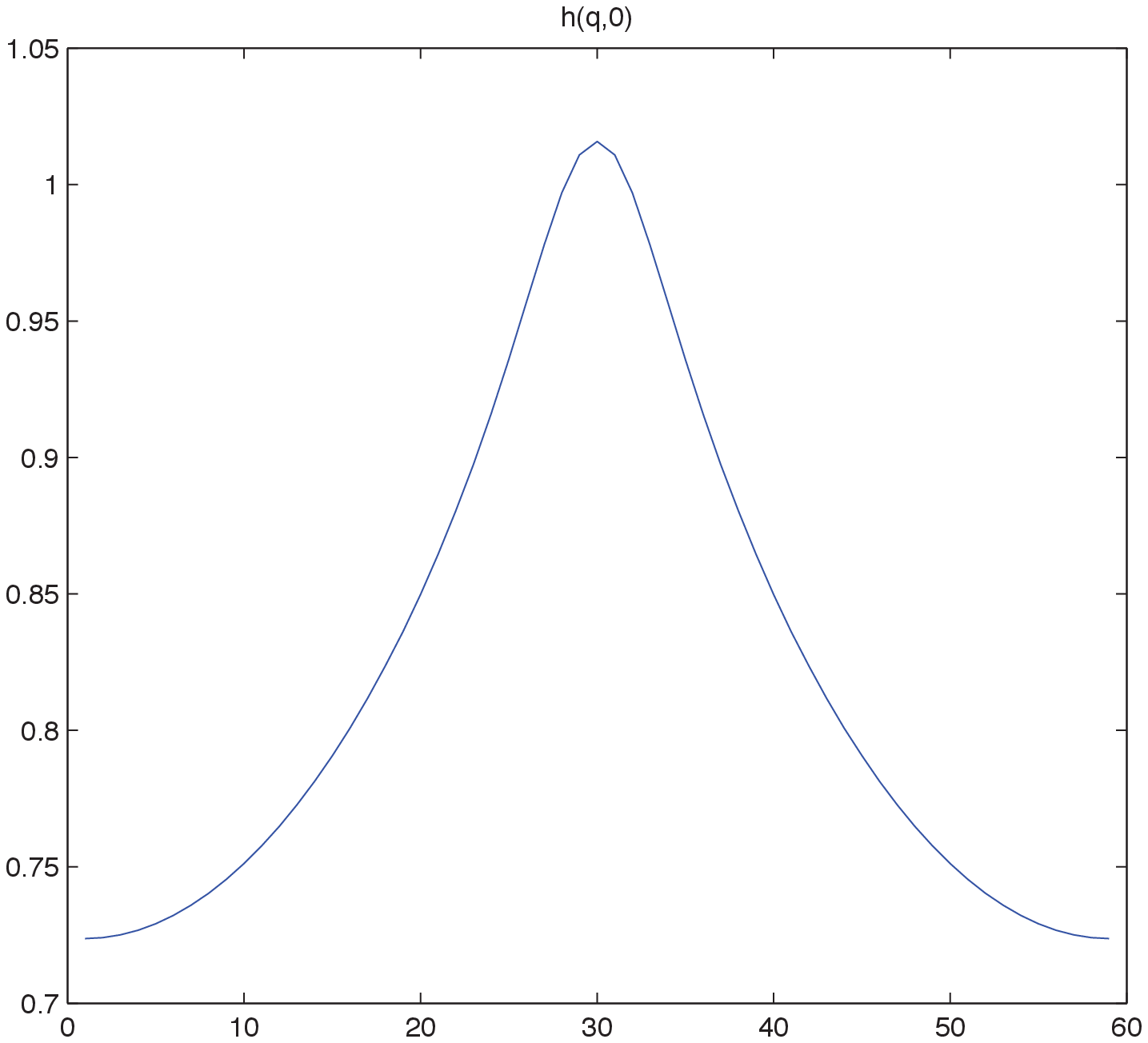}
  \caption{$\gamma = 0$}
\end{subfigure}%
\begin{subfigure}{.33\textwidth}
  \centering
  \includegraphics[width=1\linewidth]{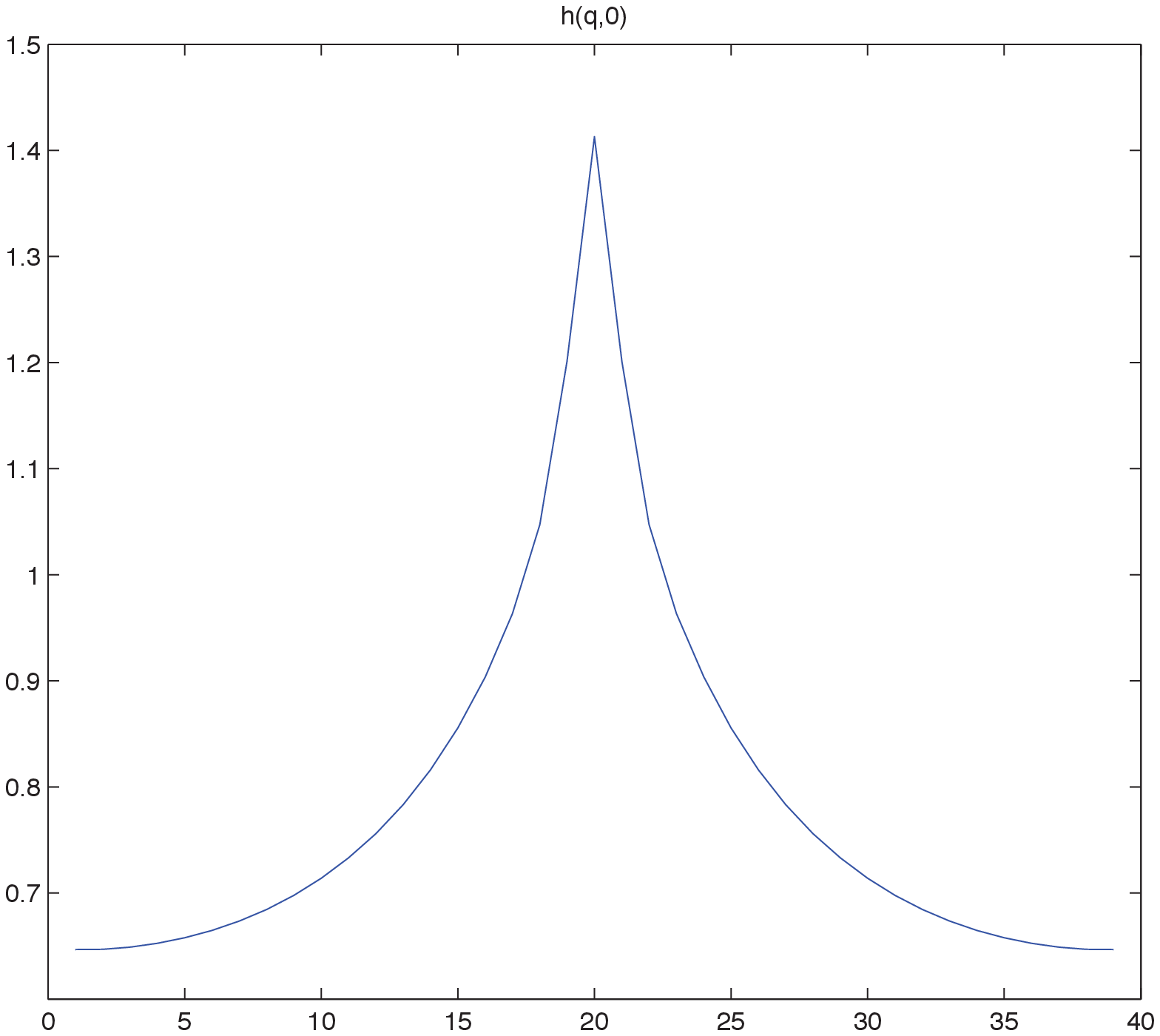}
  \caption{$\gamma = 2.95$}
\end{subfigure}
\begin{subfigure}{.33\textwidth}
  \centering
  \includegraphics[width=1\linewidth]{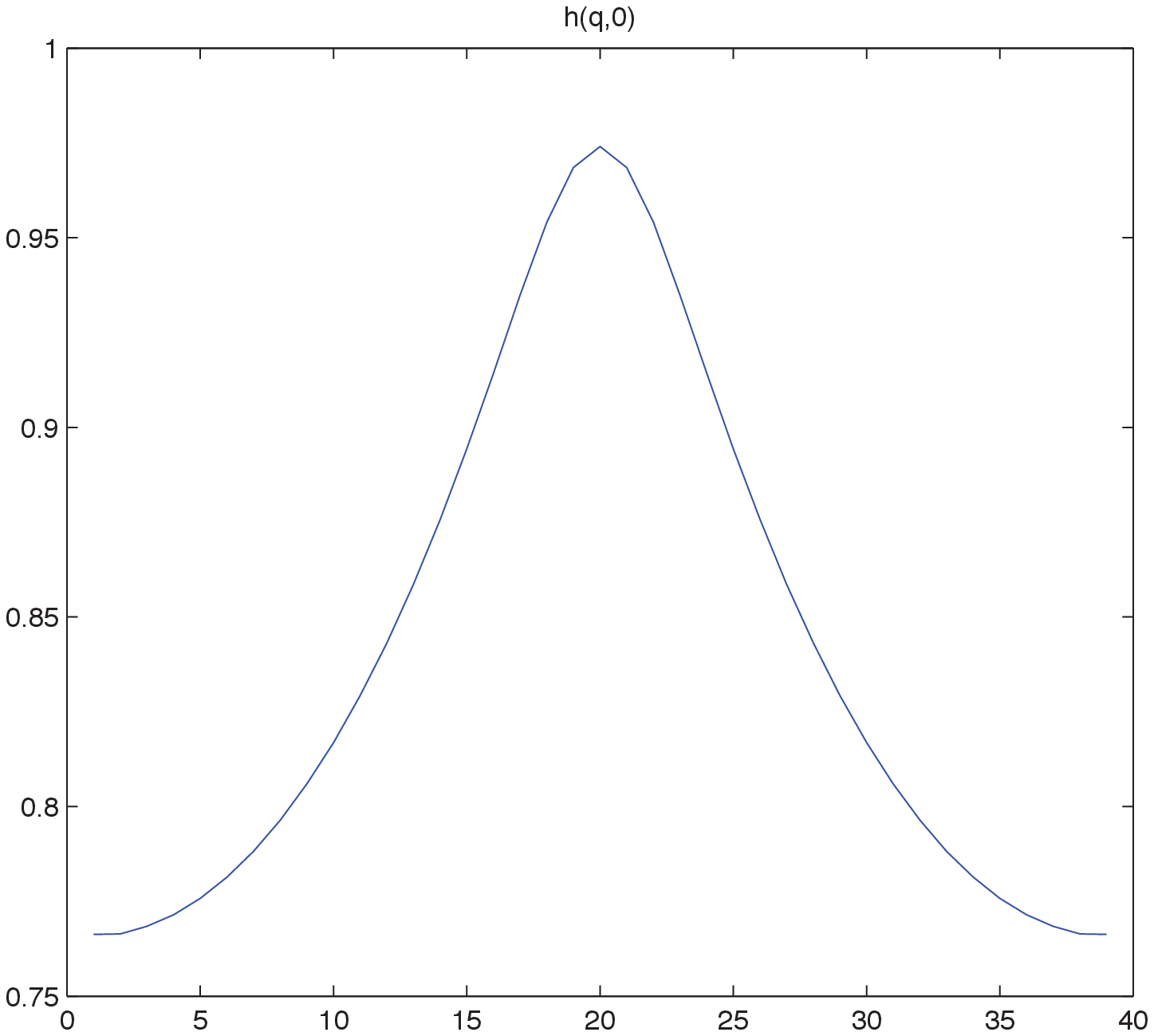}
  \caption{$\gamma = -1$}
\end{subfigure}
\caption{The periodic wave profile $S$.}
\end{figure}

\begin{figure}
\centering
\begin{subfigure}{.5\textwidth}
  \centering
  \includegraphics[width=1\linewidth]{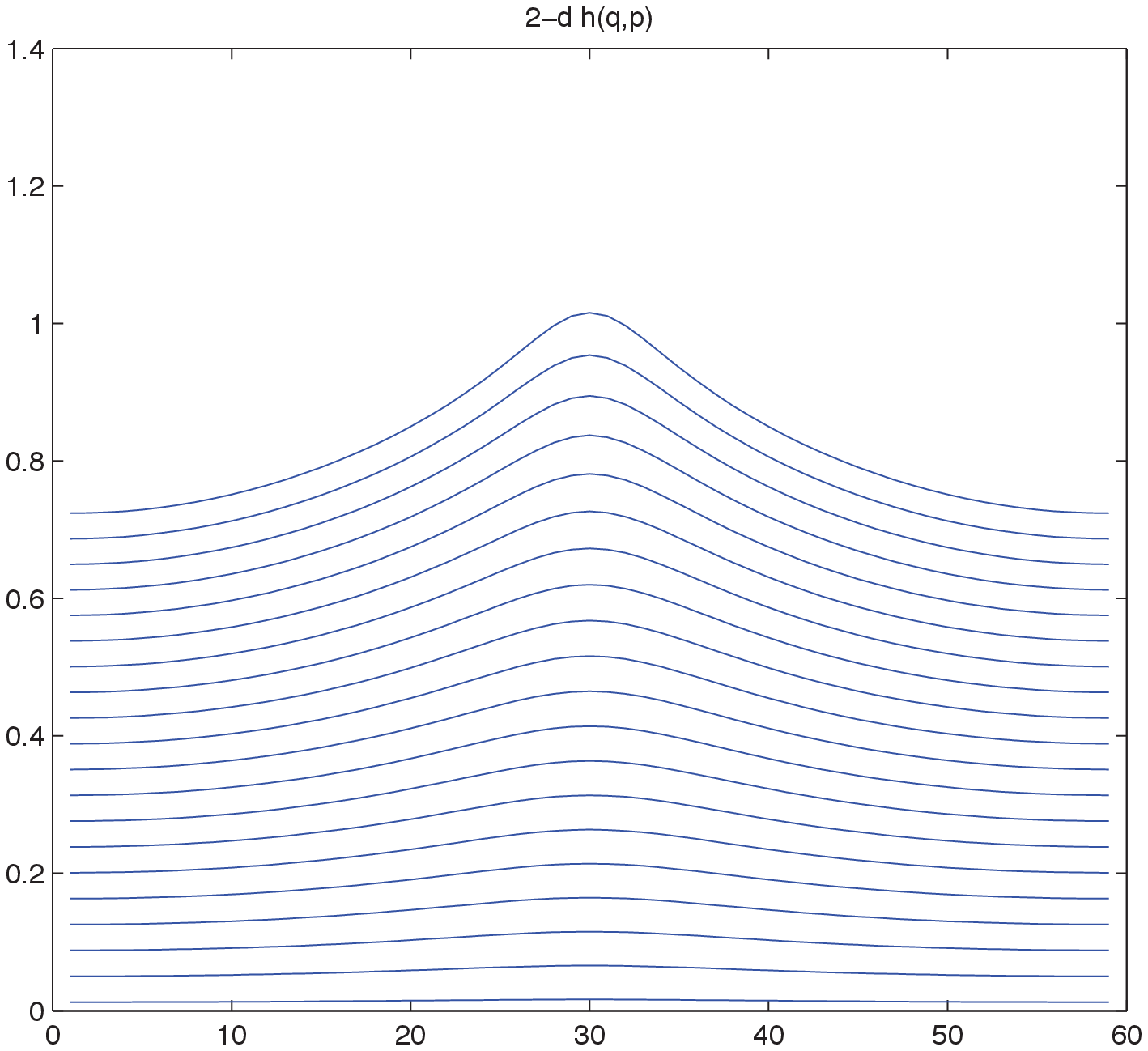}
  \caption{$h(q,p)$}
\end{subfigure}%
\begin{subfigure}{.5\textwidth}
  \centering
  \includegraphics[scale=0.42]{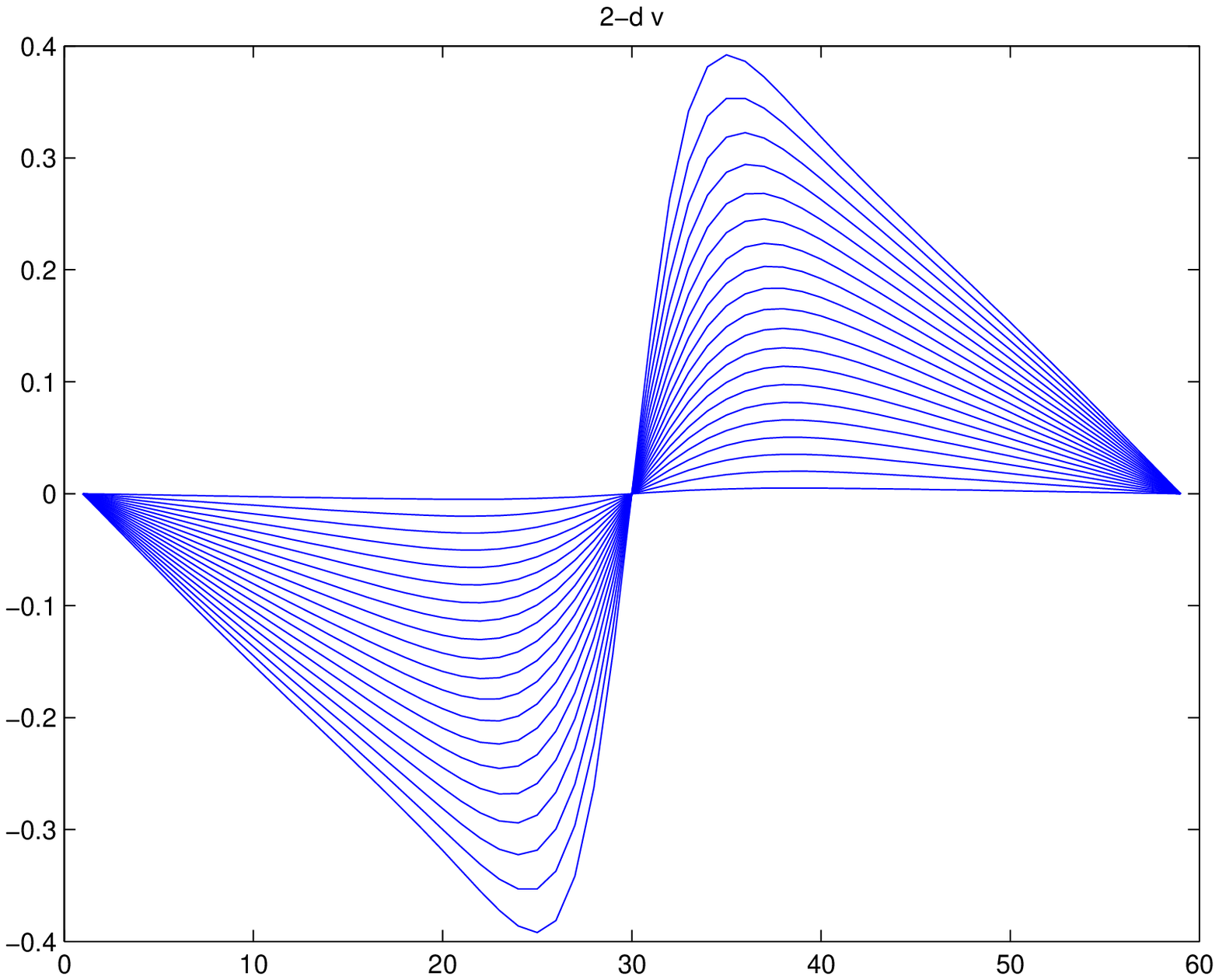} 
  \caption{$v(q,p)$}
\end{subfigure}
\caption{The height $h(q,p)$ of the streamlines and the vertical fluid velocity $v(q,p)$ along streamlines, depicted 
for the irrotational case $\gamma=0$.}
\end{figure}

\begin{figure}
\centering
\begin{subfigure}{.5\textwidth}
  \centering
  \includegraphics[width=1\linewidth]{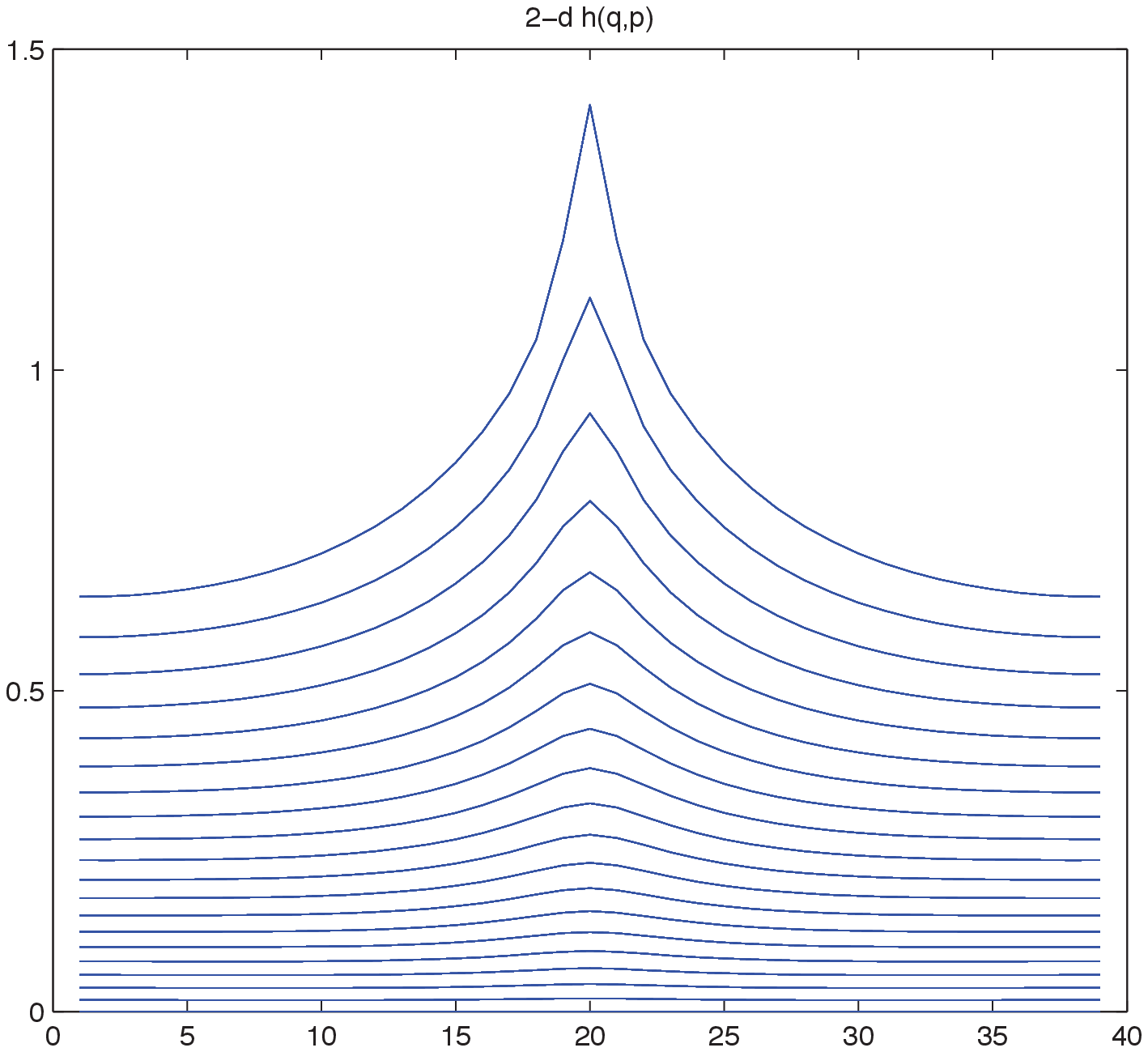}
  \caption{$h(q,p)$}
\end{subfigure}%
\begin{subfigure}{.5\textwidth}
  \centering
  \includegraphics[scale=0.42]{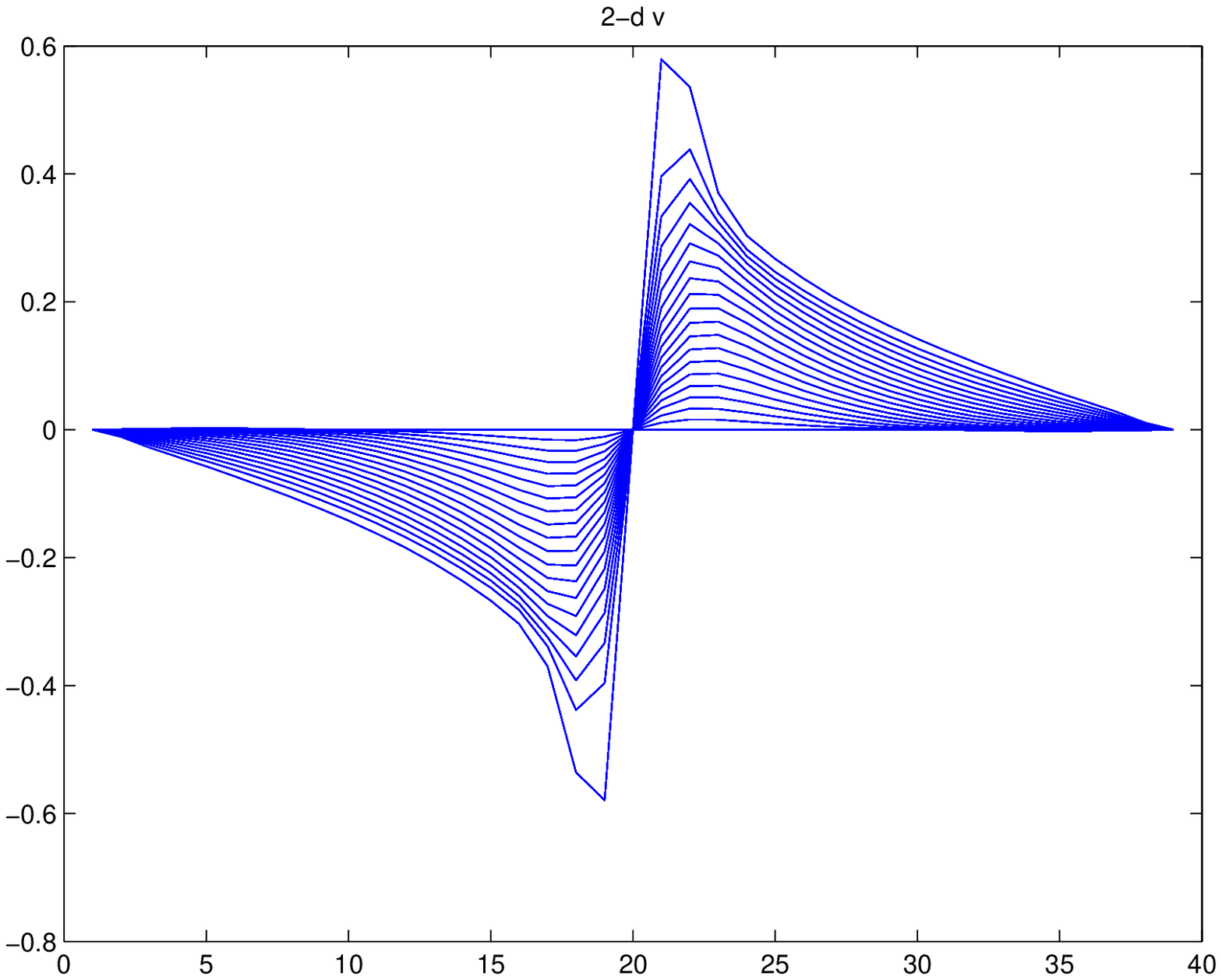} 
  \caption{$v(q,p)$}
\end{subfigure}
\caption{The height $h(q,p)$ of the streamlines and the vertical fluid velocity $v(q,p)$ along streamlines, depicted for the case $\gamma=2.95$.}
\end{figure}

\begin{figure}
\centering
\begin{subfigure}{.5\textwidth}
  \centering
  \includegraphics[width=1\linewidth]{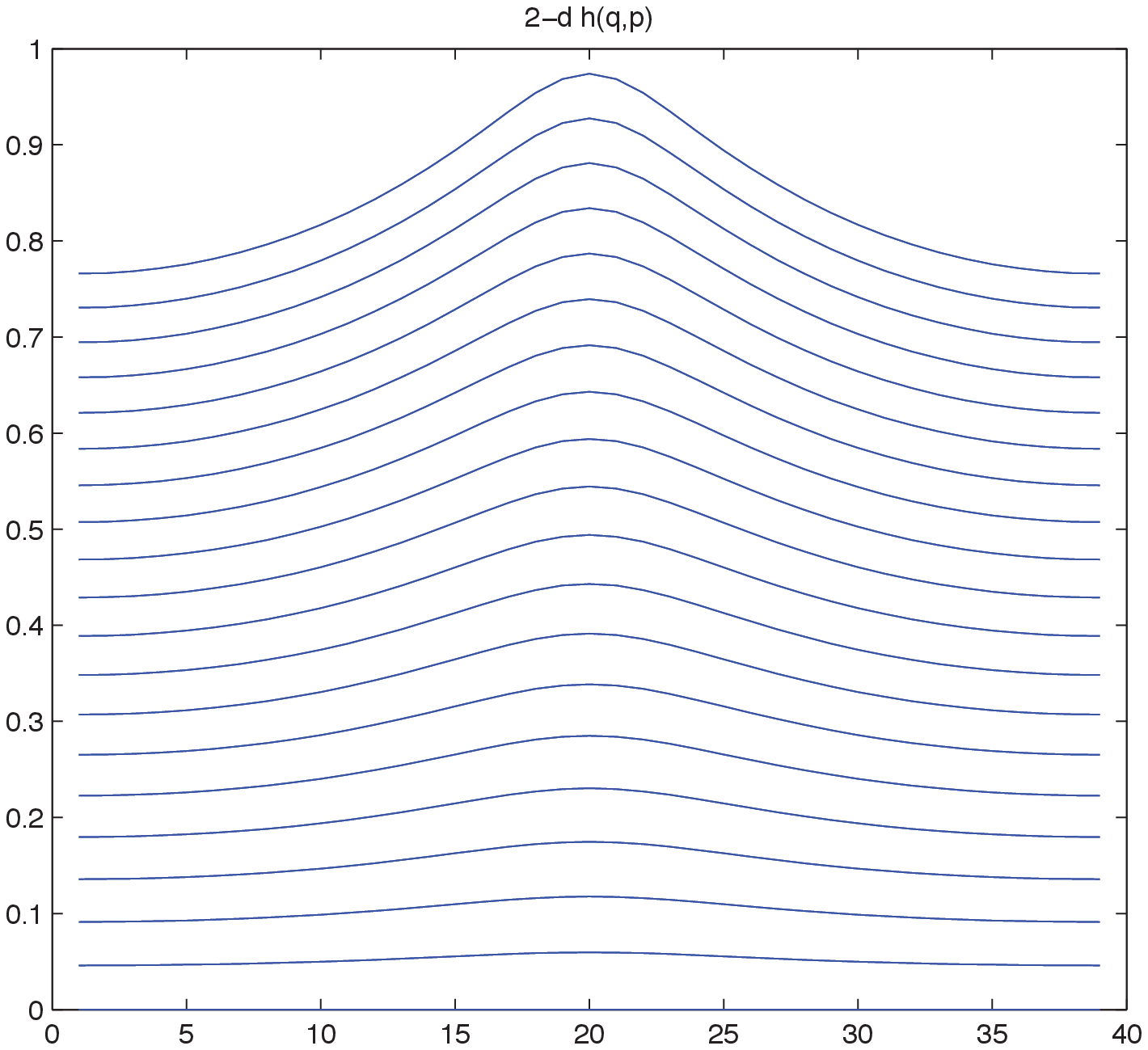}
  \caption{$h(q,p)$}
\end{subfigure}%
\begin{subfigure}{.5\textwidth}
  \centering
  \includegraphics[scale=0.42]{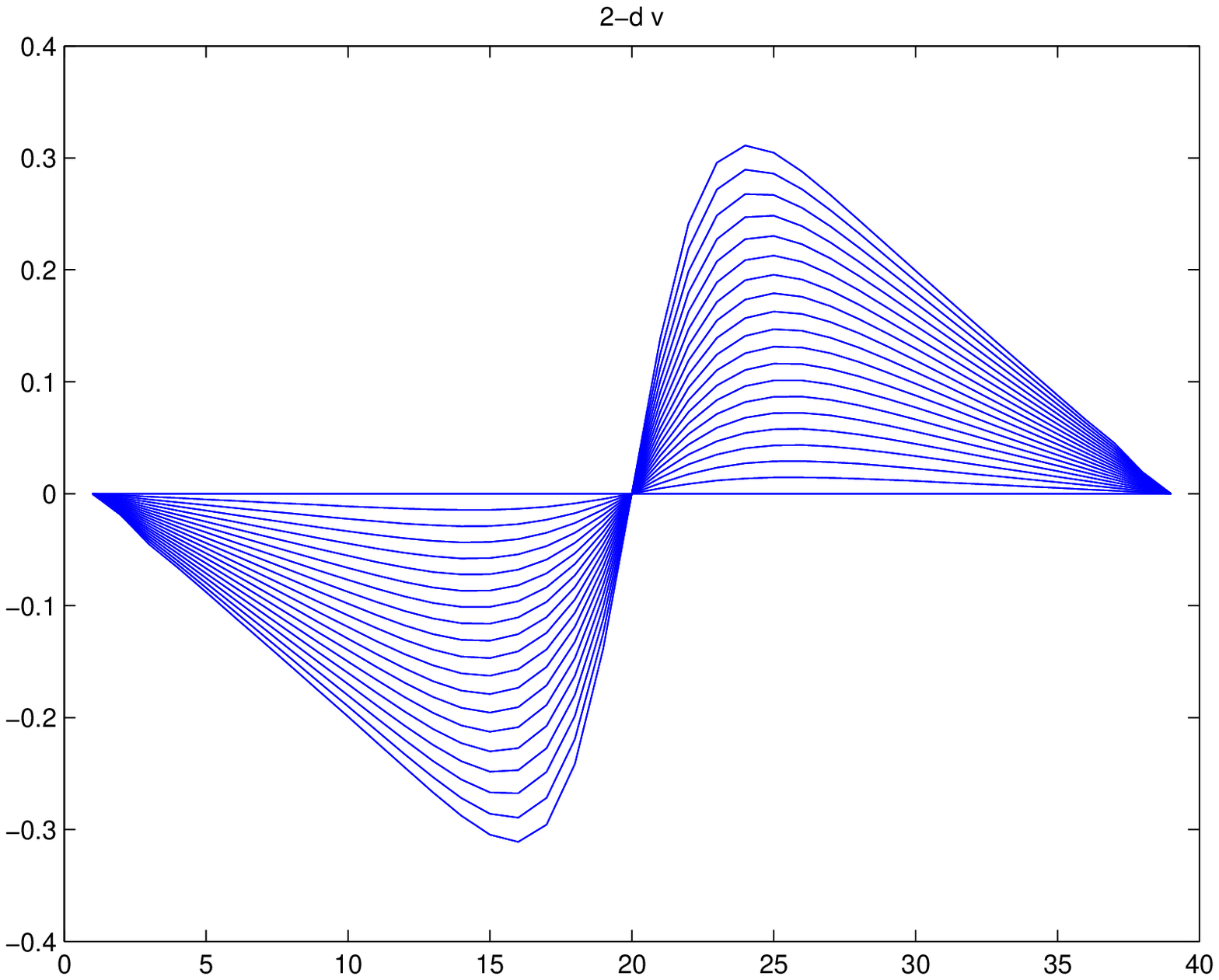} 
  \caption{$v(q,p)$}
\end{subfigure}
\caption{The height $h(q,p)$ of the streamlines and the vertical fluid velocity $v(q,p)$ along streamlines, depicted for the case $\gamma=-1$.}
\end{figure}

\begin{figure}
\centering
\begin{subfigure}{.33\textwidth}
  \centering
  \includegraphics[width=1\linewidth]{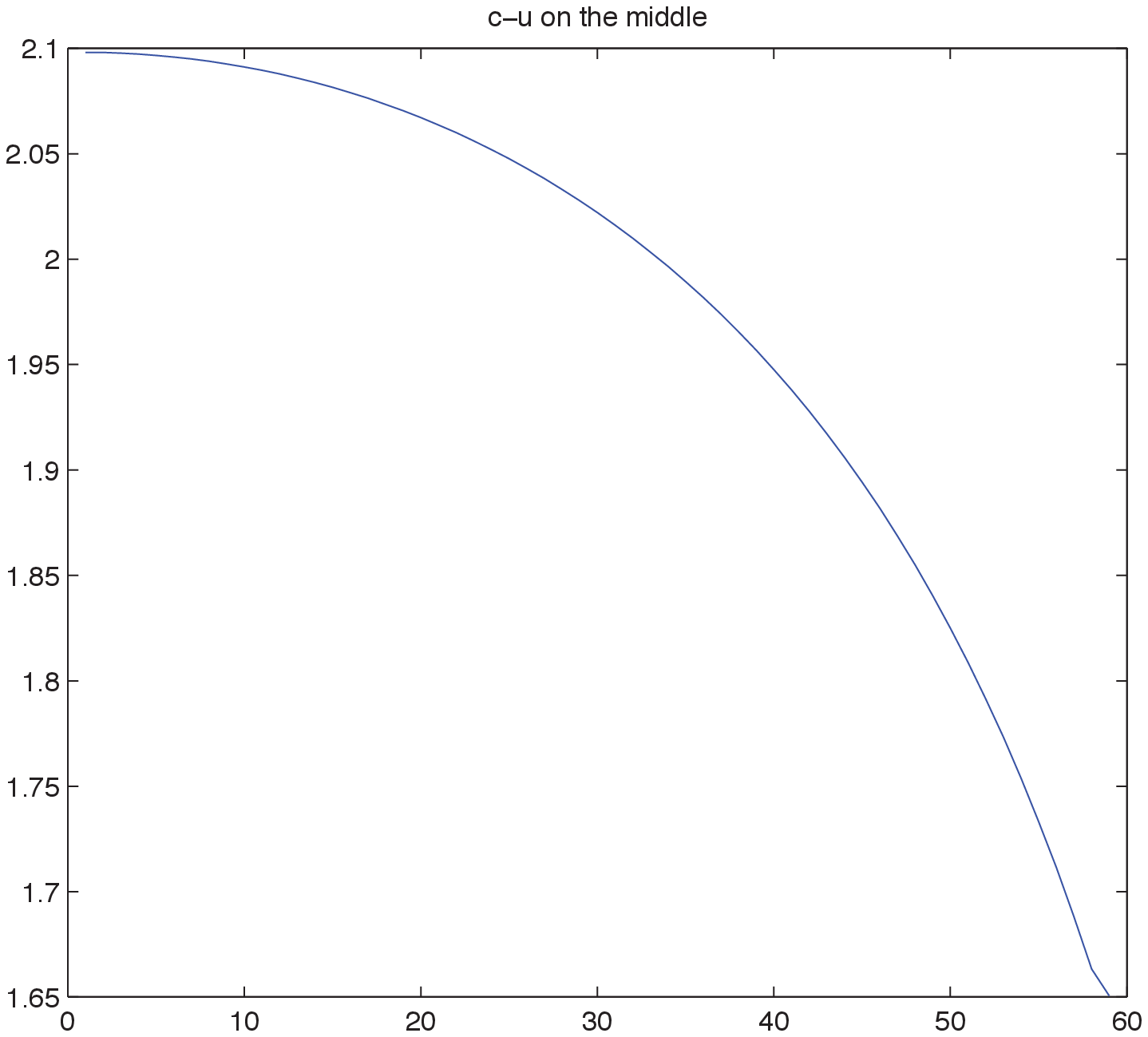}
  \caption{$\gamma = 0$}
\end{subfigure}%
\begin{subfigure}{.33\textwidth}
  \centering
  \includegraphics[width=1\linewidth]{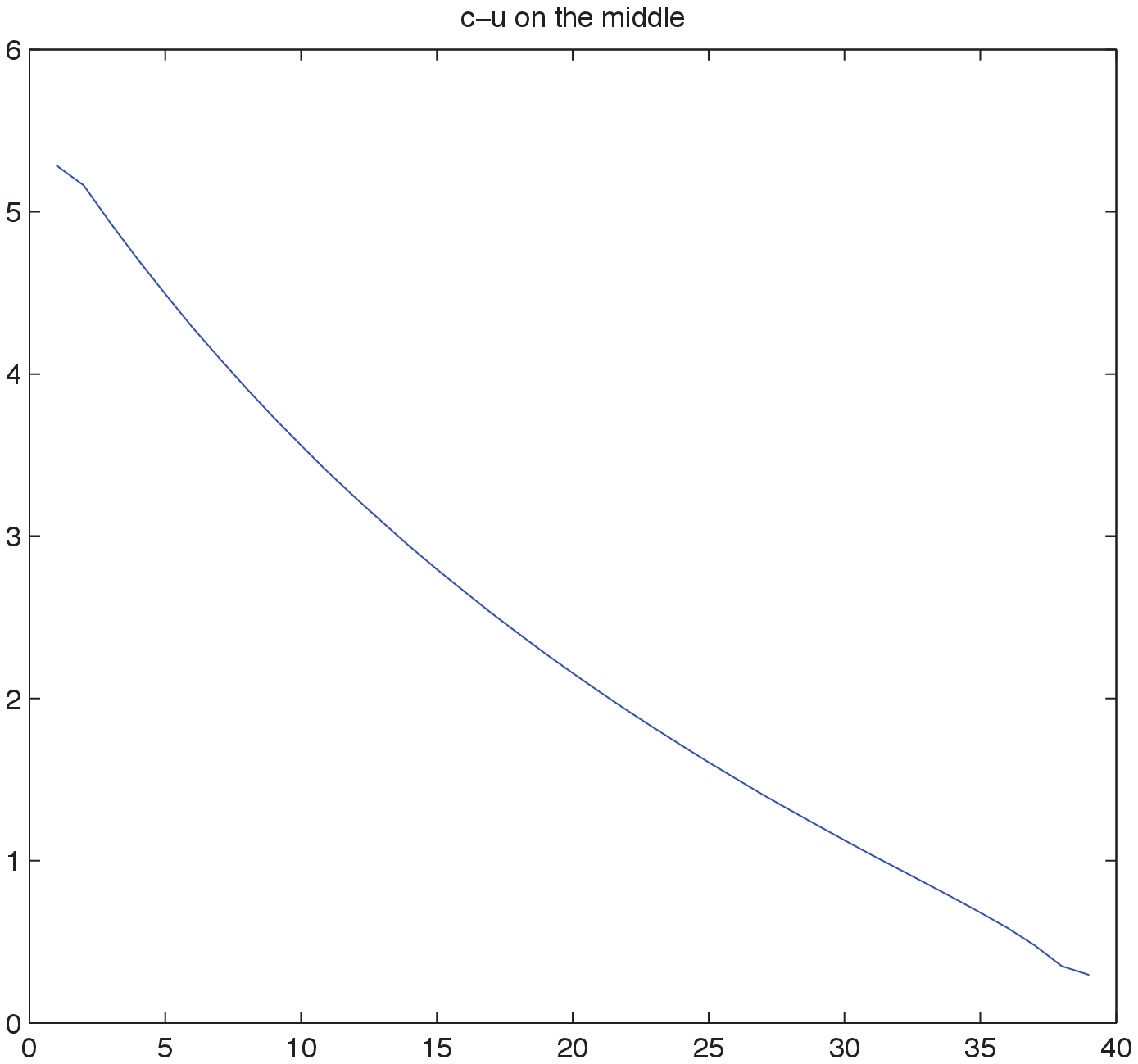}
  \caption{$\gamma = 2.95$}
\end{subfigure}
\begin{subfigure}{.33\textwidth}
  \centering
  \includegraphics[width=1\linewidth]{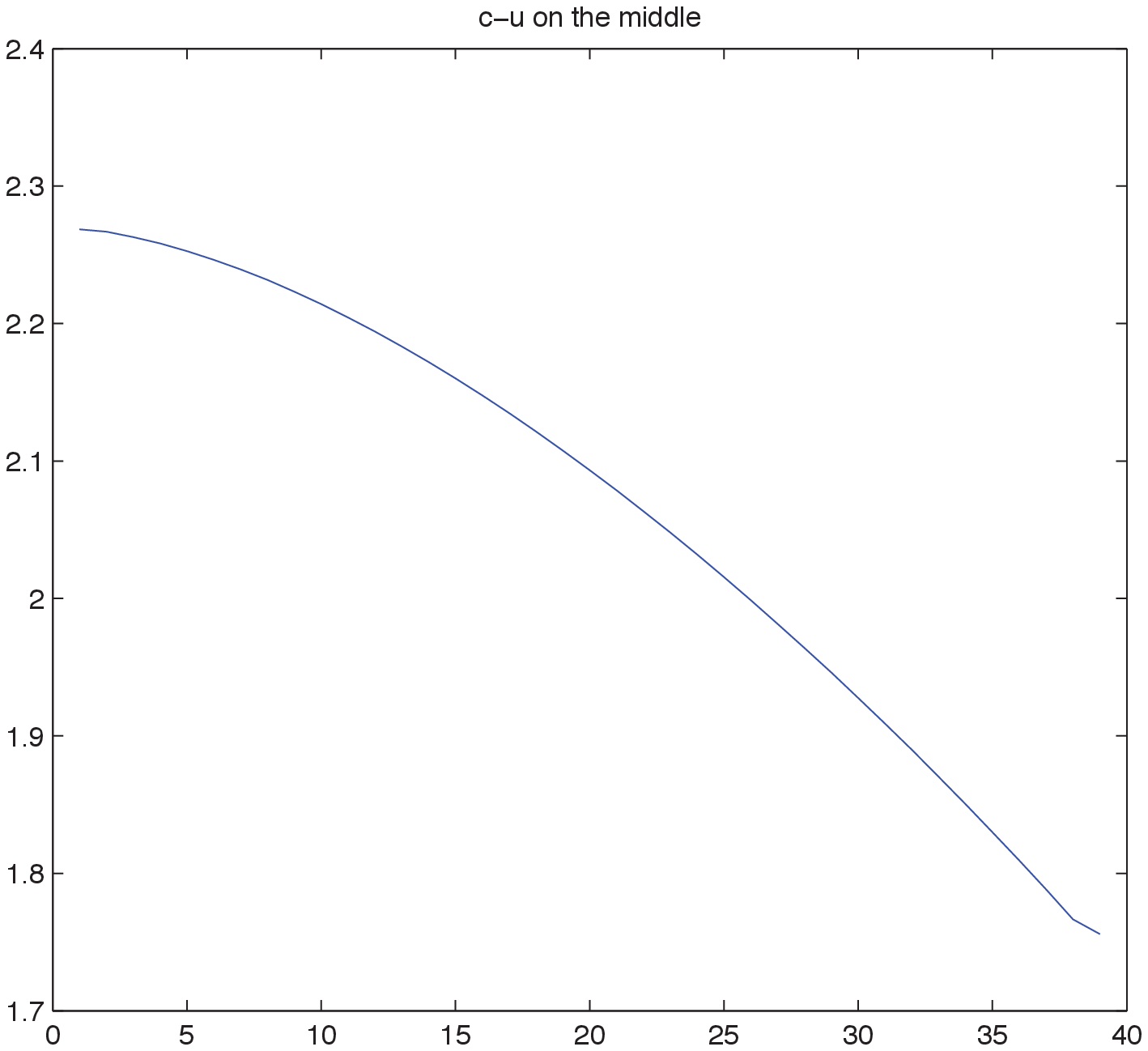}
  \caption{$\gamma = -1$}
\end{subfigure}
\caption{The horizontal fluid velocity $c-u$ beneath the wave crest, on the line segment $q=0$.}
\end{figure}

\begin{figure}
  \centering
  \includegraphics[scale=0.6]{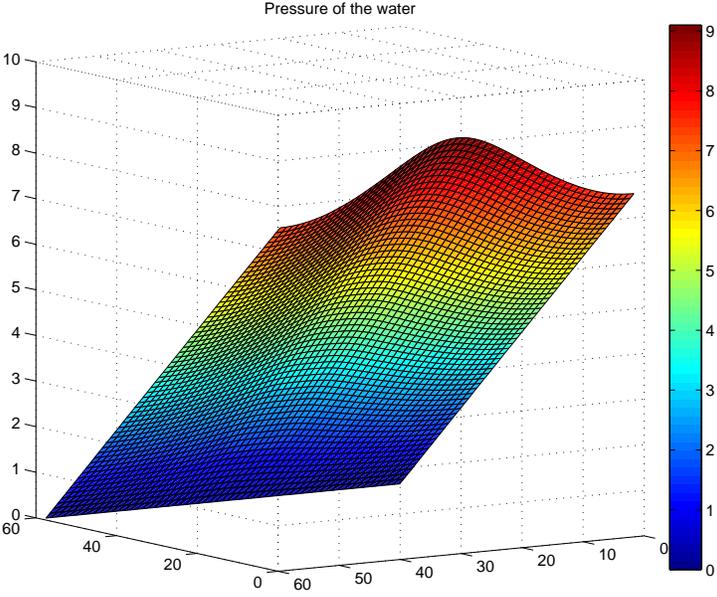}
  \caption{The pressure beneath the wave for $\gamma=0$. The figure depicts the deviation of the pressure from the (constant) 
atmospheric pressure, ${\frak p}-P_{atm}$: vanishing at the free surface $p=0$, it increases as we descend towards the flat bed $p=p_0$, the maximum 
being attained on the bed just below the wave crest (located in the middle of the horizontal segment for the discretization we
have made).}
\end{figure}%

\begin{figure}
\centering
\begin{subfigure}{.49\textwidth}
  \centering
  \includegraphics[width=1\linewidth]{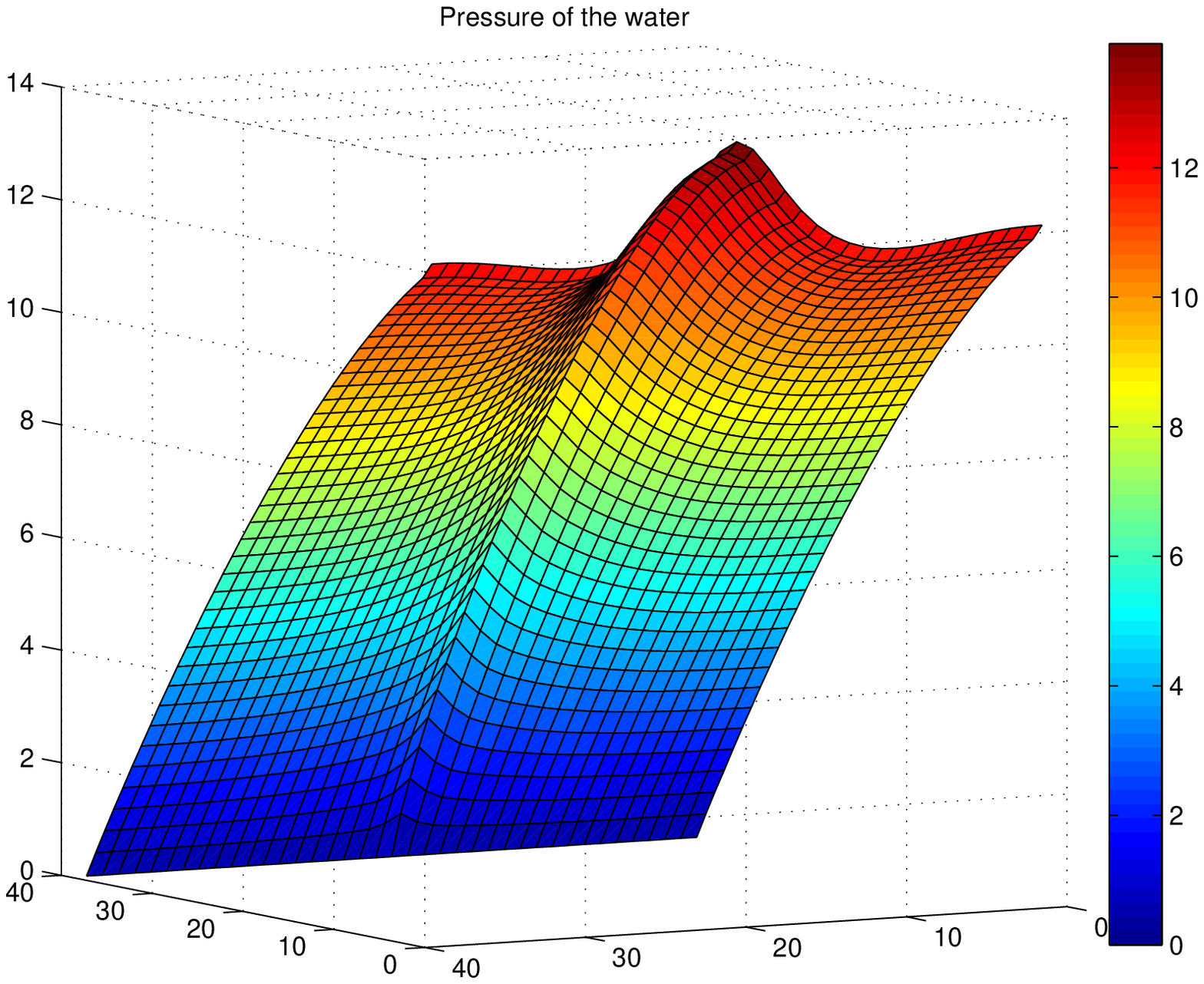}
  \caption{$\gamma = 2.95$}
\end{subfigure}
\begin{subfigure}{.49\textwidth}
  \centering
  \includegraphics[width=1\linewidth]{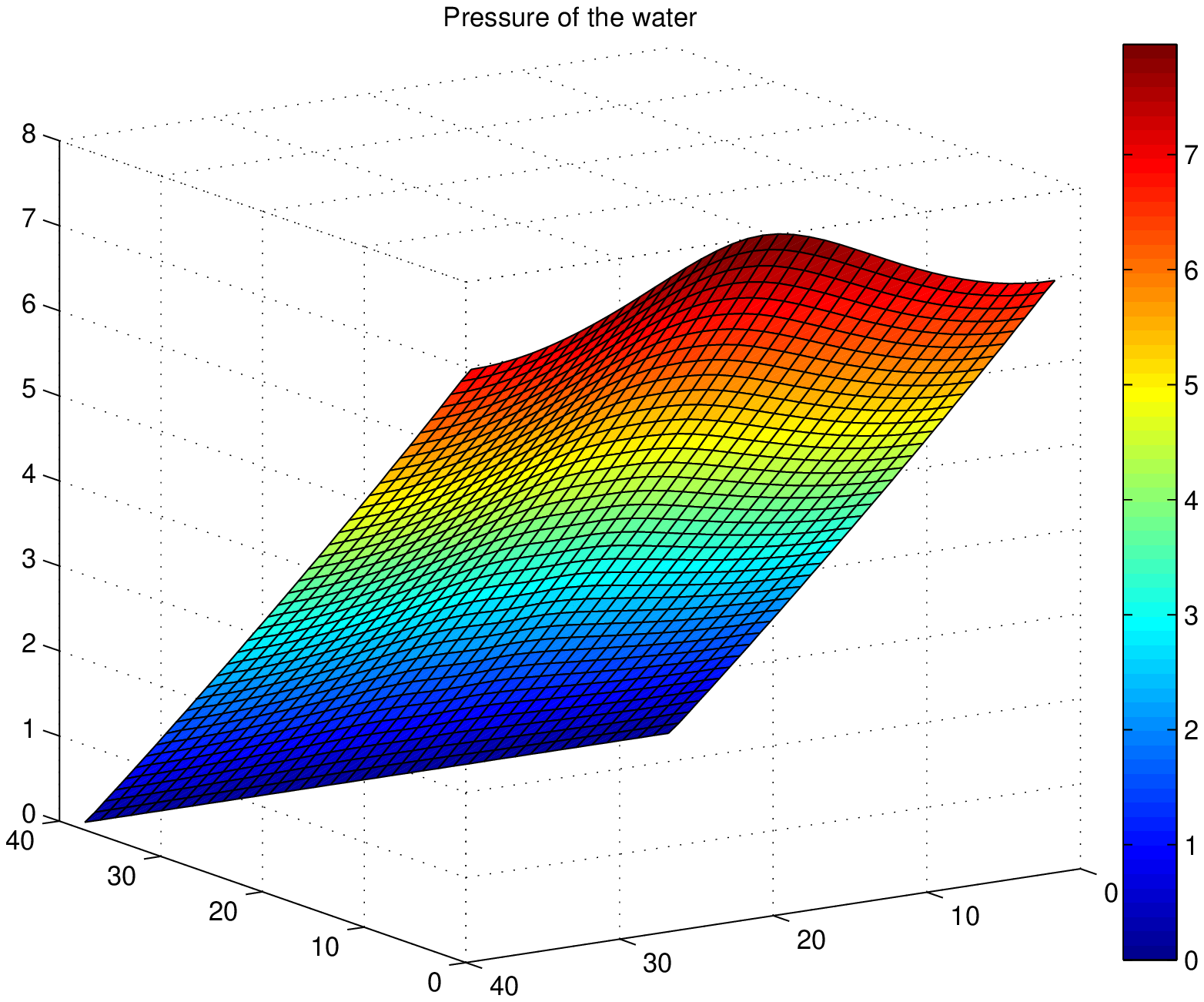}
  \caption{$\gamma = -1$}
\end{subfigure}
\caption{The pressure beneath rotational water waves.}
\end{figure}

%

\section{Conclusion}

In the present paper we analysed  a penalization method for computing two-dimensional travelling water waves. We provided an iterative algorithm that starts with an approximation of a solution of the system \eqref{op_basic} and converges to a solution which correspond to a water wave of large amplitude. The formula of the initial approximation is given by  \eqref{sol_perturb} for the non-zero vorticity case; the irrotational case is given as a special case by simply substituting $\gamma=0$. 

Moreover, our pursuit for a better approximation of a non-laminar solution (which would serve the initial step of our iterative algorithm) has lead to novel analytical results. In particular, explicit formulas that approximate non-laminar (large amplitude) travelling water waves with constant vorticity are obtained. The relevant analysis and formulas are to be presented in upcoming work.

\end{document}